\documentclass[pdflatex,sn-mathphys-num]{sn-jnl}
\usepackage{mathrsfs}
\usepackage{lmodern}
\usepackage{fix-cm}

\usepackage{tempora}
\usepackage{amsmath}

\usepackage[T2A]{fontenc}
\usepackage{lmodern}
\usepackage[russian,english]{babel}

\usepackage{graphicx}%
\usepackage[printonlyused,withpage,nolist]{acronym}

\usepackage[utf8]{inputenc}
\usepackage{multirow}%
\usepackage{mathrsfs}
\usepackage{caption}
\usepackage{amsmath,amssymb,amsfonts}%
\usepackage{amsthm}%
\usepackage{mathrsfs}%

\usepackage[utf8]{inputenc} 
\usepackage[russian, english]{babel}  
\usepackage[T2A]{fontenc}
\usepackage{lmodern} 

\usepackage{graphicx}
\usepackage{graphicx}
\usepackage{caption}
\usepackage{subcaption}
\usepackage{natbib}
\usepackage[title]{appendix}%
\usepackage{xcolor}%
\usepackage{textcomp}%
\usepackage{manyfoot}%
\usepackage{booktabs}%
\usepackage{algorithm}%
\usepackage{algorithmicx}%

\usepackage{algpseudocode}%
\usepackage{mathrsfs} 
\usepackage{listings}%
\usepackage{placeins} 

\theoremstyle{thmstyleone}%
\theoremstyle{thmstyletwo}%
\usepackage{braket}
\theoremstyle{thmstylethree}%
\DeclareUnicodeCharacter{0443}{y}
\DeclareUnicodeCharacter{041C}{M}
\DeclareUnicodeCharacter{043B}{l}
\DeclareUnicodeCharacter{0438}{i}
\DeclareUnicodeCharacter{0446}{ts}

\usepackage{float}
\begin{document}

\title{
  \centering
  Enhancement of \\[0.5em]
  Quantum Semi-Supervised Learning via \\[0.5em]
  Improved Laplacian and Poisson Methods
}


\author*[1]{\fnm{Hamed}\sur{ Gholipour}}\email{hamed.gholipour@ubi.pt}

\author[2]{\fnm{Farid} \sur{ Bozorgnia}}\email{f.bozorgnia@newuu.uz}
\equalcont{These authors contributed equally to this work.}

\author[3,4]{\fnm{Hamzeh} \sur{ Mohammadigheymasi}}\email{hmohammadigheymasi@fas.harvard.edu}
\equalcont{These authors contributed equally to this work.}

\author[1]{\fnm{Kailash}\sur{ Hambarde}}\email{kailas.srt@gmail.com}

\author[5]{\fnm{Javier}  \sur{ Mancilla}}\email{  javier@falcondale.pro}
\equalcont{These authors contributed equally to this work.}

\author[6]{\fnm{Hugo}  \sur{Proença}}\email{hugomcp@di.ubi.pt}
\equalcont{These authors contributed equally to this work.}

\author[1]{\fnm{Joao}  \sur{ Neves}}\email{jcneves@ubi.pt}
\equalcont{These authors contributed equally to this work.}

\author[7,8]{\fnm{Moharram}  \sur{Challenger}}\email{moharram.challenger@uantwerpen.be}
\equalcont{These authors contributed equally to this work.}

\affil*[1]{\orgdiv{Department of Computer Science}, \orgname{ University of Beira Interior}, \orgaddress{\street{R. Marquês de Ávila e Bolama}, \city{Covilhã}, \postcode{6201-001},\country{Portugal}}}

\affil[2]{\orgdiv{Department of Mathematics}, \orgname{New Uzbekistan University}, \orgaddress{\street{улица Мовароуннахра}, \city{Tashkent}, \postcode{1}, \country{Uzbekistan}}}

\affil[3]{\orgdiv{Faculty of Art and Sciences}, \orgname{Harvard University}, \orgaddress{\street{20 Oxford Street}, \city{Cambridge}, \postcode{02138}, \state{MA}, \country{USA}}}  

\affil[4]{\orgdiv{Atmosphere and Ocean Research Institute}, \orgname{The University of Tokyo}, \orgaddress{\street{5-1-5 Kashiwanoha}, \city{Kashiwa}, \postcode{277-8564}, \state{Chiba}, \country{Japan}}} 

\affil[5]{\orgdiv{Falcondale LCC}, \city{Delware},\country{US}}

\affil[6]{\orgdiv{Instituto de Telecomunicações}, \orgname{University of Beira Interior}, \orgaddress{\street{R. Marquês de Ávila e Bolama}, \city{Covilhã}, \postcode{6201-001}, \country{Portugal}}}  

\affil[7]{\orgdiv{Department of Computer Science}, \orgname{University of Antwerp}, \orgaddress{\city{Antwerp}, \postcode{2020}, \country{Belgium}}}

\affil[8]{\orgdiv{AnSyMo/Cosys Corelab}, \orgname{Flanders Make Strategic Research Center}, \orgaddress{\city{Leuven}, \postcode{3001}, \country{Belgium}}}


\abstract{ This paper develops a hybrid quantum approach for graph-based semi-supervised learning to enhance performance in scenarios where labeled data is scarce. We introduce two enhanced quantum models, the Improved Laplacian Quantum Semi-Supervised Learning (ILQSSL) and the Improved Poisson Quantum Semi-Supervised Learning (IPQSSL) that incorporate advanced label propagation strategies within variational quantum circuits. These models utilise QR decomposition to embed graph structure directly into quantum states, thereby enabling more effective learning in low-label settings. We validate our methods across four benchmark datasets—\textit{Iris}, \textit{Wine}, \textit{Heart Disease}, and \textit{German Credit Card} and show that both ILQSSL and IPQSSL consistently outperform leading classical semi-supervised learning algorithms, particularly under limited supervision. Beyond standard performance metrics, we examine the effect of circuit depth and qubit count on learning quality by analyzing entanglement entropy and Randomized Benchmarking (RB). Our results suggest that while some level of entanglement improves the model’s ability to generalize, increased circuit complexity may introduce noise that undermines performance on current quantum hardware. Overall, the study highlights the potential of quantum-enhanced models for semi-supervised learning, offering practical insights into how quantum circuits can be designed to balance expressivity and stability. These findings support the role of quantum machine learning in advancing data-efficient classification, especially in applications constrained by label availability and hardware limitations.}




\maketitle

\section{Introduction}\label{sec1}

\ac{SSL} has become a crucial method in machine learning, addressing the prevalent scenario where a large amount of unlabeled data is readily available, but labeled data is scarce and costly to obtain~\cite {van2020survey}. \ac{SSL} has found widespread applications in various fields, such as computer vision~\cite{xie2020self}, biomedical data analysis~\cite{zhu2003semi}, seismology~\cite{mohammadigheymasi2023ipiml}, and natural language processing~\cite{miyato2016adversarial}, by efficiently propagating label information into unlabeled datasets. The ability of \ac{GSSL} to model data as graphs sets it apart from other \ac{SSL} approaches. This ability naturally fits with the manifold assumption that high-dimensional data frequently resides on a lower-dimensional structure. \ac{GSSL} allows label propagation throughout the dataset by representing data points as nodes and their similarities as weighted edges. This effectively captures the underlying data topology as well as the limited supervision provided by the few labeled samples, enabling more accurate classification or regression in sparse label scenarios.

A major challenge in \ac{GSSL} methods is the effective propagation of the labels to unlabeled data, particularly when the labeled samples are extremely scarce. For example, Laplacian learning~\cite{zhu2003semi}, a widely used method in manifold classification~\cite{yang2013saliency}, propagates label information across the dataset through the solution of a graph-based harmonic function. However, because of limited robustness in label propagation, it often struggles to achieve high accuracy, particularly when supervision is minimal. The \emph{p}-Laplace extension of the Laplacian framework attempts to improve robustness by incorporating a parameter \( p \) into the diffusion process. When \( p > 2 \), it extends the conventional Laplacian formulation, and in the extreme case of \( p = \infty \), it corresponds to the Lipschitz learning~\cite{kyng2015algorithms, bozorgnia2024graphbasedsemisupervisedsegregatedlipschitz}. Although \emph{p} Laplace learning improves label propagation, its non-linear nature significantly increases computational complexity~\cite{el2016asymptotic}. More recently, Poisson learning has emerged as a promising alternative, demonstrating strong classification performance even at extremely low label rates~\cite{calder2020poisson}. 

At the same time, advances in quantum computing have opened up new areas for improving graph-based learning. Quantum computing can utilize principles such as superposition and entanglement, offering computational advantages that could surpass classical methods \cite{beer2023quantum}. This has led to the rise of Quantum Graph Learning (QGL), where graph structures are encoded into quantum states, enabling new ways to process and analyze data using quantum mechanics\cite{Cerezo_2022}. Several studies have explored how quantum computing can increase traditional graph-based learning, improving scalability and efficiency \cite{tang2022quantumgraphcomputingquantum}. Among these approaches, Laplacian-based Quantum Semi-Supervised Learning has been studied in depth, highlighting how tuning system parameters can significantly impact classification performance \cite{gholipour2025laplacian}.

In this work, we build on these ideas by implementing an advanced hyperparameter tuning strategy to improve Laplacian Quantum Semi-Supervised learning (ILQSSL), improving both label propagation and classification accuracy. Additionally, we incorporate Improved Poisson Quantum Semi-Supervised Learning (IPQSSL), and we implement experiments on four well-known benchmark datasets: Iris \cite{misc_iris_53}, Wine \cite{misc_wine_109}, German Credit Card \cite{statlog_(german_credit_data)_144}, and Heart Disease \cite{misc_heart_disease_45}. Beyond standard classification metrics, we systematically analyze the effects of key quantum properties, such as entanglement and RB, by varying the number of qubits and layers in quantum circuits. This enables us to investigate how quantum effects impact graph-based learning from both theoretical and experimental perspectives.

The structure of this paper is as follows. \textbf{Section 2} introduces the fundamental concepts of GL and QGL, highlighting their theoretical foundations. \textbf{Section 3} explores the application of the ILQSSL method to four distinct datasets. In the same section, we also describe both the Improved Poisson Learning (IPL) and IPQSSL approaches, with evaluation of the accuracy of IPQSSL in comparison with the classical methods in the previously mentioned data sets. \textbf{Section 4}investigates the role of entanglement and performs RB on parametrized quantum circuits, examining the effects of varying circuit depth and qubit counts. Lastly, the Section 7 ROC/AUC and the Kolmogorov-Smirnov(KS) statistics for our data sets are investigated.
In \textbf{Section 5}, the principal findings of the study are examined and interpreted in detail. The discussion covers their significance, considers existing limitations, and proposes areas for future exploration. The concluding part summarizes the study’s contributions, emphasizes the importance of the results, and identifies promising directions for continued research.

\section{Background}\label{sec2}
\subsection{Graph and Quantum Graph Learning }
GL is a branch of machine learning that focuses on analysing data graphs to identify trends and derive pertinent information. A graph is made up of nodes that represent entities and edges that display their relationships \cite {skolik2023equivariant}. Knowledge graphs, social networks, molecular structures, and automated transportation systems are examples of real-world applications for this structure \cite{luo2022automated}. One of the main issues with GL is the difficulty of processing graph data efficiently, particularly when resource constraints are present. The adjacency list, which connects each node to its neighborsusing a list; the adjacency matrix, which is a grid with node connections indicated; and the incidence matrix, which shows the relationships between nodes and edges through a binary matrix, are common representations.

Although GL has advanced considerably, there are still some key limitations that reduce its effectiveness. One of the main issues is its inability to capture long-range information. Traditional GL methods often struggle to gather data over long distances within a graph. As graphs become larger, the information is condensed into fixed vectors, which limits the flow of data between distant nodes, especially when interactions with distant data are crucial \cite{yu2022graph}. Another challenge is the lack of explainability and interpretability in many graph models. These models can be difficult to understand, making it difficult to interpret how decisions are made or explain the output of the model in a meaningful way. Although interpretable models offer clear human-understandable explanations, explainable models allow for a more thorough understanding of the reasoning of the model after the fact \cite{miao2022interpretable}. Finally, working with large-scale graphs, like those representing brain networks, can be overwhelming for classical graph learning techniques. To overcome this, approaches such as breaking the graph into smaller subgraphs or using parallel processing methods for learning are often explored.

Quantum Graph Learning (QGL) can be broadly categorized into three main areas: Quantum Computing on Graphs, Quantum Graph Representation, and Quantum Graph Neural Networks (QGNNS). The emergence of Noisy Intermediate-Scale Quantum (NISQ) devices has opened new avenues for the development of QGNNs. These hybrid models integrate classical Graph Neural Networks (GNNS) with quantum computing techniques to improve learning performance. Two primary strategies have been proposed for realizing QGNNs: (1) Quantum Algorithms for Classical GNNS and (2) Quantum Circuits for Modifying GNN Architectures. The first strategy employs fault-tolerant quantum algorithms to accelerate computations traditionally handled by classical GNNs. The second approach introduces parametrized quantum circuits (PQCs) composed of fixed and tunable quantum gates that approximate the objective function by adjusting the gate parameters during training.

\begin{figure*}[t]
    \centering
    \includegraphics[width=\textwidth]{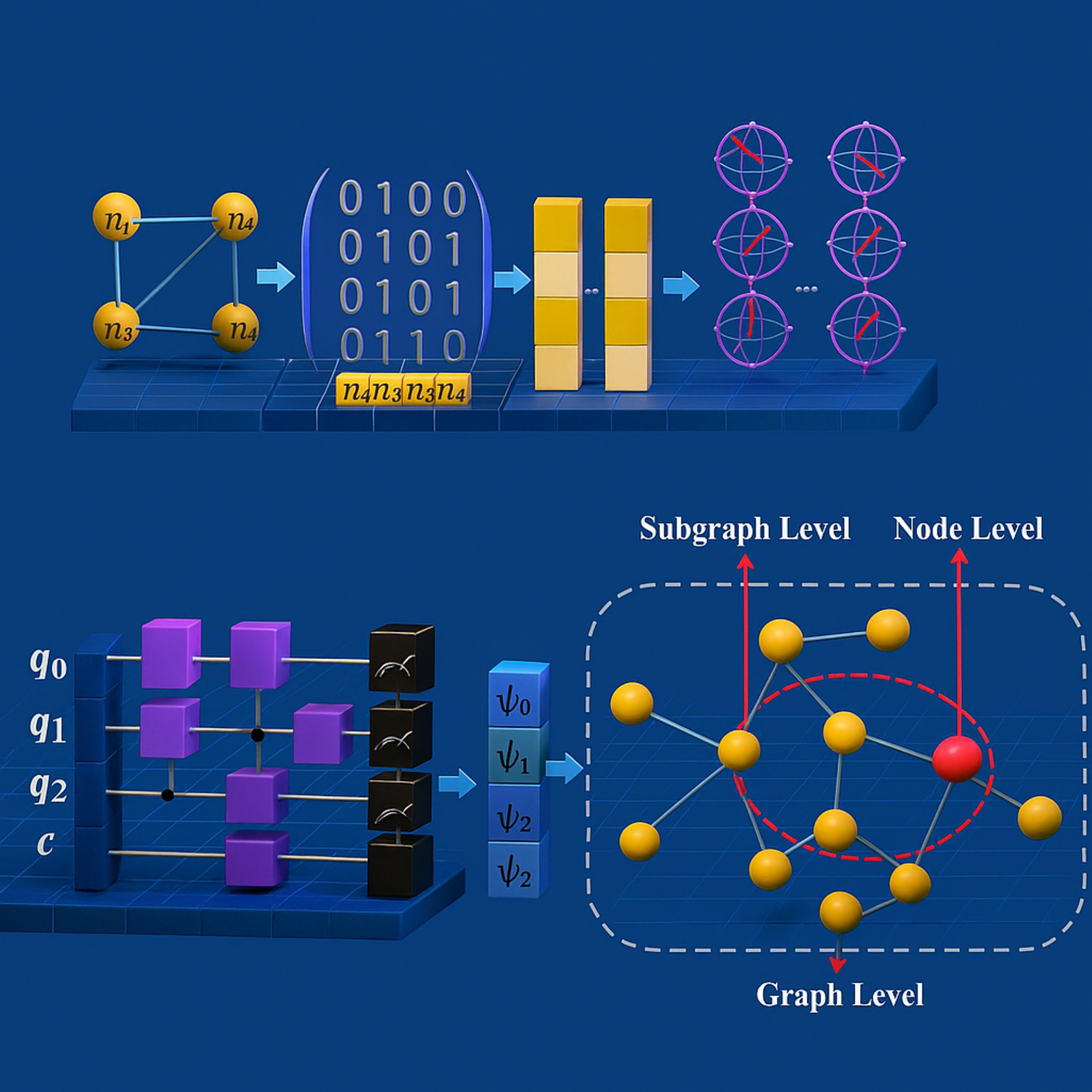}
    \caption{Quantum circuit architecture used in this study, illustrating key components such as entangling layers, unitary operations, and quantum gates~\cite{yu2023quantumgraphlearningfrontiers}.}
    \label{fig:example}
\end{figure*}

\subsubsection{Limitations in Quantum Graph Learning}

Although QGL offers significant advantages, several challenges remain. The complexity of quantum mechanics can be counterintuitive, and while a basic understanding of linear algebra and complex numbers is sufficient for (QGL), mastering the nuances of quantum systems is challenging. The practical constraints of quantum hardware present another limitation, as quantum devices currently have limited qubit numbers, restricting their ability to process large-scale graph data. However, hybrid quantum-classical methods are being explored to bridge this gap. 
In addition, the incompatibility with the classical frameworks poses difficulties. Although quantum circuits can offer speed-ups in GL, integrating quantum and classical components in hybrid models can introduce challenges to achieve full compatibility\cite{Bharti_2022}.

\subsubsection{Quantum Implementation}

Initially, in semi-supervised QGL, an adjacency matrix representing the structure of a given graph is constructed. QR decomposition uniquely embeds this adjacency matrix within parameterized quantum circuits, thus directly encoding the structural information of the graph in the quantum model \cite{Schuld:2019bfg}. This innovative method combines classical graph theory principles with quantum computation to improve classification accuracy beyond traditional variational quantum methods.

Both Laplacian and Poisson matrices are crucial in graph-based semi-supervised learning, and each can be distinctly encoded within quantum states using amplitude encoding. In amplitude encoding, the quantum state \(|\psi\rangle\) encapsulates graph data through amplitudes as:

\[ |\psi\rangle = \sum_i \beta_i |i\rangle \]

Each amplitude \(\beta_i\) corresponds directly to node \(i\) within the graph.

For Laplacian-based learning, quantum algorithms such as Quantum Phase Estimation (QPE) and the Harrow–Hassidim–Lloyd (HHL) algorithm are utilized \cite{Abrams_1999}. These quantum algorithms efficiently handle core linear algebra tasks, including solving linear equations and eigenvalue problems involving the Laplacian matrix \(L\). The HHL algorithm, for example, addresses equations of the form \(Lx = b\), first encoding the vector \(b\) into a quantum state, then applying quantum operations to derive the quantum representation of the solution \(x\). Practically, the adjacency matrix \(A\), degree matrix \(D\), and resultant Laplacian matrix \(L\) are encoded into quantum states. Quantum algorithms subsequently process these states to compute the solution vector \(f\), facilitating smooth label propagation across the graph structure.
For Poisson-based learning, a similar quantum computational approach is adopted. Quantum algorithms again leverage QPE and HHL, solving Poisson equations that emerge naturally from graph structures. Specifically, these algorithms tackle linear systems characterized by Poisson matrices, facilitating efficient propagation of label information\cite{BravoPrieto2023variationalquantum}. Quantum states representing the Poisson matrices are encoded through amplitude encoding, enabling quantum circuits to exploit the inherent structure of the data for effective classification and prediction tasks.
By clearly delineating the mathematical foundations and quantum implementation methodologies, both Laplacian and Poisson methods offer robust frameworks for graph-based semi-supervised learning\cite{skolik2023equivariant}. These quantum-enhanced techniques use intrinsic graph structures, generalizing known labels to unlabeled data points effectively and potentially outperforming classical graph-based methods by harnessing quantum computational advantages(Figure 1).

\section{Enhancement of Quantum Graph-Based Semi-Supervised Learning}
\subsection{Improved Laplacian Quantum Graph-based Semi-Supervised Learning}\label{sec3}

In the paper \cite{gholipour2025laplacian} Laplacian-based Quantum-Graph Neural Networks approach was explained across four benchmark datasets, such as Iris, Wine, Breast Cancer Wisconsin, and Heart Disease. We examine how varying qubit counts and entangling layers affect performance. Our analysis suggests that increasing quantum resources does not always improve outcomes. The method's effectiveness is closely tied to the unique characteristics of each dataset and the number of entangling layers. Generally, optimal performance is achieved with moderate entanglement, which balances model complexity and the ability to generalize. These results underline the need for dataset-specific adjustments when tuning hyperparameters in quantum semi-supervised learning approaches.
This section focuses on the use of hyperparameter tuning as an optimization technique for Laplacian-based Quantum Graph Neural Networks (QGNNS), applied across datasets: Iris, Wine, and Heart Disease. Hyperparameter tuning refers to the process of identifying the most effective combination of model parameters, such as the number of layers, qubits, and graph neighbors, to enhance the model’s performance on a specific task.

\subsection {Analysis of Model Performance of Improved Laplacian Quantum Semi-Supervised Learning(ILQSSL)}

\begin{table}[htbp]
\centering
\label{tab:comparison_lqssl}
\begin{tabular}{|p{3cm}|p{3cm}|c|c|c|c|}
\hline
\textbf{Dataset} & \textbf{Method} & \textbf{Test Accuracy} & \textbf{F1} & \textbf{Recall} & \textbf{Precision} \\
\hline
\multirow{2}{*}{Iris} & Improved LQSSL & 0.95 & 0.93 & 1.00 & 0.88 \\
                      & Original LQSSL & 0.822 & 0.779 & 0.938 & 0.666 \\
\hline
\multirow{2}{*}{Wine} & Improved LQSSL & 0.65 & 0.83 & 0.38 & 0.53 \\
                      & Original LQSSL & 0.533 & 0.515 & 0.753 & 0.391 \\

\hline
\multirow{2}{*}{Heart Disease} & Improved LQSSL & 0.51 & 0.40 & 0.37 & 0.47 \\
                               & Original LQSSL & 0.48 & 0.47 & 0.50 & 0.45 \\
\hline
\multirow{2}{*}{German Credit} & Improved LQSSL & 0.60 & 0.28 & 0.23 & 0.25 \\
                               & Original LQSSL & 0.59 & 0.28 & 0.23 & 0.26 \\                               
\hline
\end{tabular}
\caption{Comparison of classification metrics (Test Accuracy, F1, Recall, and Precision) between Improved and Original Laplacian Quantum Semi-Supervised Learning (LQSSL) on multiple datasets \cite{gholipour2025laplacian}}
\end{table}

Table 1 presents a comparative evaluation of the Improved and Original Laplacian Quantum Semi-Supervised Learning (LQSSL) methods across four benchmark datasets: Iris, Wine, Heart Disease, and German Credit. The Improved LQSSL demonstrates a marked performance enhancement on the Iris dataset, achieving a significant increase in test accuracy (95\% vs.\ 82.2\%) alongside superior F1 score, recall, and precision metrics. On the Wine dataset, the improved method yields a higher accuracy and F1 score but exhibits a trade-off with reduced recall and increased precision, indicating a more conservative classification behavior. For the Heart Disease and German Credit datasets, performance differences between the two methods are marginal, with the original method marginally outperforming in some recall-related metrics for Heart Disease, and negligible differences observed for German Credit. These results highlight the efficacy of the Improved LQSSL in well-structured datasets while underscoring challenges in achieving consistent gains on more complex or imbalanced data. The findings motivate further refinement of quantum semi-supervised learning algorithms tailored to diverse real-world datasets.

\subsection {Improved Poisson Learning (IPL) for Quantum Graph-based Semi-Supervised Learning}

Next, Poisson Learning, an effective technique for graph data, is applied within quantum graph learning frameworks to optimize classification accuracy, entanglement, and coherence. The improved Poisson Learning (IPL) method\cite{bozorgnia2024improved} utilizes an iterative process to optimize data class distributions and reduce time complexity. The algorithm is designed to be more efficient and capable of handling larger, more complex graphs.

To enhance the convergence and stability of Poisson Learning, we implement the Improved Poisson Learning (IPL) algorithm. The primary improvement in IPL focuses on overcoming the convergence limitations of the original iterative scheme. IPL guarantees faster convergence and better alignment with the graph’s intrinsic structure and connectivity by introducing additional regularization terms and adjusting the structure of the iterative updates, improving its performance in graph-based semi-supervised learning tasks.

 To extend our quantum algorithm, we consider the following algorithm presented in \cite{bozorgnia2024improved}. Let  $\pi=(\frac{d_1}{d}, \frac{d_2}{d} \cdots ,\frac{d_n}{d}),$  where  $d_i=\sum_j W_{ij}$, $d=\sum_i d_i$ .   Then set matrix $Q$ to be a rank one matrix with rows $\pi$.

\begin{algorithm}[H]
\caption{: Improved Poisson learning}
\label{alg:ipl}
\begin{algorithmic}[1]
\Require 
    Similarity matrix $W \in \mathbb{R}^{n \times n}$, 
    initial label matrix $Y \in \mathbb{R}^{n \times k}$, 
    parameters $\alpha_1, \alpha_2, \alpha_3 \in \mathbb{R}$, 
    tolerance $\epsilon > 0$
\Ensure 
    Predicted labels $y_i$ or final label matrix $U \in \mathbb{R}^{n \times k}$

\State \textbf{Step 1:  Compute  matrices $D = \text{diag}(d_i)$  and   Transition Matrix   $P = D^{-1} W$,   $Q$ }
  
\Statex
\State \textbf{Step 2: Initialization}
\State Initialize label matrix $U^{(0)} = D^{-1} Y$
\State Set $m = 0$

\Statex
\State \textbf{Step 3: Iterative Update}
\Repeat
    \State $U^{(m+1)} = (P - \alpha_1 Q + \alpha_2 I) U^{(m)} + \alpha_3 D^{-1} B^\top$
    \State $m = m + 1$
\Until{$\| U^{(m)} - U^{(m-1)} \|_F < \epsilon$}

\Statex
\State \textbf{Step 4: Final Converged Solution}
\State $U^* = U^{(m)}$

\Statex
\State \textbf{Step 5: Label Assignment}
\For{each data point $x_i$}
    \State $y_i = \arg\max_j U^*_{ij}$
\EndFor
\end{algorithmic}
\end{algorithm}

By subtracting \( Q \) from \( P \), IPL avoids the divergence issues present in the original Poisson algorithm. The condition   $ \underset{m \to \infty}{\lim} (P - \alpha_1 Q)^m = 0,$ ensures convergence to a stable solution. Adding the term \( \alpha_2I  \) boosts self-transition probabilities, stabilizing the iterative process and ensuring that nodes retain influence from previous states. This contributes to the faster stabilization of the learning process.  

The parameterized structure of IPL allows fine-tuning for different graph structures and datasets, making it adaptable to diverse scenarios, including those with imbalanced or sparse graph data. Additionally, the integration of the class distribution \( b \) via \( B^T \) ensures that updates respect underlying class proportions, leading to more accurate predictions, particularly in semi-supervised learning tasks with class imbalance\cite{bozorgnia2024improved}.

\begin{algorithm}[H]
\caption{Quantum-Enhanced Label Propagation (QELP)}
\begin{algorithmic}[1]
\Require Similarity matrix $W \in \mathbb{R}^{n \times n}$, label matrix $Y \in \mathbb{R}^{n \times k}$, parameters $\alpha_1$, $\alpha_2$, $\alpha_3$, tolerance $\varepsilon$
\Ensure Predicted labels $y_i$ or label matrix $U \in \mathbb{R}^{n \times k}$

\State \textbf{Step 1: Compute Graph Matrices}
\State $D \gets \mathrm{diag}(d_i),\ d_i = \sum_j W_{ij}$
\State $P \gets D^{-1} W$

\State \textbf{Step 2: Initialization}
\State $U^{(0)} \gets D^{-1} Y$
\State $m \gets 0$

\State \textbf{Step 3: Quantum Graph Encoding}
\State Perform QR decomposition: $W = U_{\text{graph}} R$
\State Encode similarity matrix: $\ket{\psi_W} = \sum_i \beta_i \ket{i}$

\State \textbf{Step 4: Iterative Update}
\Repeat
    \State Estimate $\hat{U}_{\text{iter}} \ket{\psi_{U^{(m)}}}$
    \State $U^{(m+1)} \gets (P - \alpha_1 Q + \alpha_2 I) U^{(m)} + \alpha_3 D^{-1} B^\top$
    \State $m \gets m + 1$
\Until{$\|U^{(m+1)} - U^{(m)}\|_F < \varepsilon$}

\State \textbf{Step 5: Final Converged Solution}
\State $U^* \gets U^{(m)}$

\State \textbf{Step 6: Label Assignment}
\For{each data point $x_i$}
    \State $y_i \gets \arg\max_j U^*_{ij}$
\EndFor

\State \textbf{Step 7: Quantum Evaluation}
\State Analyze observables from $\ket{\psi_{U^{(m)}}}$

\end{algorithmic}
\end{algorithm}

The Quantum-Enhanced Label Propagation (QELP) method combines classical graph-based semi-supervised learning with quantum state encoding to improve classification performance on structured datasets. The algorithm begins by constructing graph matrices from a similarity matrix \( W \in \mathbb{R}^{n \times n} \), where each element \( W_{ij} \) quantifies the similarity between nodes \( i \) and \( j \). From \( W \), the degree matrix \( D \) is defined as a diagonal matrix with \( D_{ii} = \sum_j W_{ij} \), representing the total weight of edges connected to node \( i \). The transition matrix \( P = D^{-1} W \) is then computed to model label diffusion across the graph. A residual matrix \( Q \) is also introduced to enforce label smoothness.

Label propagation is initialized by setting the label matrix as \( U^{(0)} = D^{-1} Y \), where \( Y \in \mathbb{R}^{n \times k} \) contains the initial labels: labeled nodes are assigned one-hot vectors, and unlabeled nodes are initialized with zero vectors. The degree normalization \( D^{-1} \) ensures balanced influence based on node connectivity. An iteration index \( m = 0 \) is used to track updates. Until now, all the steps have been like the classical IPL Algorithm. Before propagation begins, the graph structure is mapped into a quantum representation. This is done using QR decomposition of the similarity matrix. QR decomposition is a matrix factorization technique that decomposes a matrix into the product of an orthogonal matrix (Q) and an upper triangular matrix (R). Here, I will explain the Quantum Embedding of Graph data via QR decomposition.

\subsection{Quantum Embedding of Graph Data via QR Decomposition}

In quantum machine learning (QML), particularly for tasks such as node classification, graph classification, and quantum-enhanced message passing, embedding classical graph data into quantum states is a crucial step. Unlike classical computing, quantum systems require that all transformations be unitary, thus posing constraints on how classical information can be represented in quantum circuits. To address this challenge, QR decomposition provides a principled and hardware-compatible method for transforming classical graph-derived matrices into unitary operators suitable for quantum state preparation.

\subsubsection{QR-Based Quantum Embedding}

Given a graph \( G = (V, E) \), each node \( v_i \in V \) is associated with a feature vector \( x_i \in \mathbb{R}^d \). To embed this into a quantum state, a data-dependent matrix \( A_i \in \mathbb{R}^{d \times d} \) is constructed, which may include features from node \( v_i \) and its neighbors. Applying QR decomposition yields:

\[
A_i = Q_i R_i,
\]

where \( Q_i \) is an orthonormal matrix and \( R_i \) is upper triangular. As only unitary matrices are allowed in quantum circuits, we discard \( R_i \) and use \( Q_i \) as the unitary operator for state preparation. By applying \( Q_i \) to the initial state \( |0\rangle^{\otimes d} \), we obtain:

\[
|\phi_i\rangle = Q_i |0\rangle.
\]

This quantum state captures both the node's features and local graph structure. Applying this procedure to all nodes generates a dataset of quantum-encoded node states, which can be input into quantum classifiers, kernel methods, or variational circuits.

\subsubsection{Extensions to Graph Classification and Message Passing}

This QR-based embedding strategy generalizes to various settings. For graph classification, features of all nodes can be concatenated into a global matrix, and its QR decomposition provides a unitary that encodes the graph as a whole. In node classification, each node’s features are embedded independently using its respective \( Q_i \), allowing a shared quantum classifier to operate on node-level states.

For message-passing architectures, inspired by classical graph neural networks (GNNs), the messages exchanged between nodes can also be encoded using QR decomposition. Each message matrix is decomposed, and the orthonormal component is used to define the corresponding quantum gate operations. This forms a modular quantum analog of classical message passing, compatible with quantum hardware.

\subsubsection{Label Propagation in the Quantum Domain}

Quantum-enhanced label propagation can be integrated with the QR embeddings. The propagation rule is expressed as:

\[
U^{(m+1)} = (P - \alpha_1 Q + \alpha_2 I)U^{(m)} + \alpha_3 D^{-1} B^\top,
\]

where:
\begin{itemize}
    \item \( P \) is a similarity matrix,
    \item \( Q \) captures structural graph information,
    \item \( I \) is the identity matrix,
    \item \( D^{-1} B^\top \) introduces supervised labels,
    \item \( \alpha_1, \alpha_2, \alpha_3 \) are hyperparameters.
\end{itemize}

The updated label matrix \( U^{(m+1)} \) may be encoded as a quantum state \( |\psi_{U^{(m+1)}}\rangle \) and evolved using a unitary operator \( \hat{U}_{\text{iter}} \). This quantum evolution continues until convergence, determined by:

\[
\|U^{(m)} - U^{(m-1)}\|_F < \varepsilon,
\]

where \( \|\cdot\|_F \) denotes the Frobenius norm and \( \varepsilon \) is a small threshold. After convergence, predictions are made via:

\[
y_i = \arg\max_j U^*_{ij},
\]

where \( U^* = U^{(m)} \) is the final label matrix.

\subsubsection{Interpretability and Quantum Insights}

One of the unique advantages of the quantum approach is the ability to analyze quantum properties during the learning process. Metrics such as entanglement entropy, fidelity between quantum states, and observable expectation values can be computed. These quantities help interpret how well the model captures data structure and label propagation dynamics. They can also be used for debugging and tuning quantum models.


QR decomposition offers a robust and theoretically sound method for embedding graph data into quantum states. It satisfies the unitarity requirement of quantum circuits while leveraging classical structure. This technique supports scalable implementations, from node-level embeddings to full-graph representations and quantum message passing. Furthermore, its compatibility with quantum hardware—via decomposition into elementary gates or high-level abstractions like \texttt{QubitUnitary}—makes it a practical tool for quantum graph learning models. When integrated with iterative label propagation and quantum observables analysis, QR-based embeddings form a complete framework for advancing quantum-assisted semi-supervised learning.

\subsection{Improved Poisson Quantum semi-supervised Learning (IPQSSL)}

This section takes a closer look at classification parameters, with a focus on accuracy. The findings show that these methods perform better in terms of accuracy than the traditional ones.

\FloatBarrier

\begin{table}[h]
\centering
\begin{tabular}{p{6cm}cccc}
\toprule
\textbf{Dataset} & \textbf{Test Accuracy} & \textbf{F1} & \textbf{Recall} & \textbf{Precision} \\
\midrule
Iris & 0.97 & 0.96 & 0.96 & 0.96 \\
Wine & 0.94 & 0.94 & 0.94 & 0.95 \\
German Credit Card & 0.75 & 0.71 & 0.74 & 0.72 \\
Heart Disease & 0.83 & 0.77 & 0.70 & 0.85 \\
\bottomrule
\end{tabular}
\caption{Classification results using IPQSSL across four datasets}\label{tmethods_datasets}

\end{table}

\FloatBarrier
\begin{table}[htpb]
\centering
\begin{tabular}{lccc}
\toprule
\textbf{Dataset} & \textbf{Best Model} & \textbf{Accuracy} \\
\midrule
Iris & Label Propagation / Label Spreading & 0.911111 \\
Wine & Self-Training (SVM) & 0.722222 \\
Heart Disease & Label Propagation / Label Spreading / Self-Training (SVM) & 0.533333 \\
German Credit & Self-Training (SVM) & 0.710000 \\
\bottomrule
\end{tabular}
\caption{Highest Accuracy per Dataset}
\label{tab:best_accuracy}
\end{table}

\begin{table}[htbp]
    \centering
    \begin{tabular}{p{0.3\textwidth}p{0.3\textwidth}p{0.3\textwidth}}
        \toprule
        \textbf{Dataset} & \textbf{Best Classical Accuracy (\%)} & \textbf{IPQSSL Accuracy (\%)} \\
        \midrule
        Iris & 0.91 & 0.97 \\
        Wine & 0.72 & 0.94 \\
        Heart Disease & 0.53 & 0.83\\
        German Credit Card & 0.71 & 0.77 \\
        \bottomrule
    \end{tabular}
    \caption{Comparison of IPQSSL accuracy versus the best-performing classical model across four datasets. IPQSSL consistently outperforms classical approaches.}     \label{tab:ipl_vs_classical}
\end{table}

Tables 2, 3, and 4 collectively illustrate the superior performance of the proposed IPQSSL (Improved Quantum Semi-Supervised Learning) framework across multiple datasets compared to leading classical semi-supervised methods. In Table \ref{tmethods_datasets}, IPQSSL consistently demonstrates strong predictive performance, achieving test accuracies of 97 on Iris, 94 on Wine, 83 on Heart Disease, and 75 on the German Credit Card datasets. Notably, the associated F1-scores, recall, and precision metrics are also uniformly high, indicating balanced performance across classes and suggesting robustness against class imbalance issues often prevalent in semi-supervised settings. Table \ref{tab:best_accuracy} provides the context for these results by reporting the best-performing classical semi-supervised models for each dataset. Classical methods such as Label Propagation, Label Spreading, and Self-Training with SVMs achieved significantly lower accuracies, ranging from 53.3 (Heart Disease) to 91.1 (Iris). This disparity is particularly evident on more challenging datasets like Heart Disease, where classical methods fail to generalize effectively. A direct comparison in Table \ref{tab:ipl_vs_classical} reinforces IPQSSL's advantage: it outperforms the best classical accuracy by margins of 6 (Iris), 22 (Wine), 30 (Heart Disease), and 6 (German Credit Card), respectively. These consistent improvements across datasets of varying sizes, feature complexities, and label distributions underscore the adaptability and generalization capability of the IPQSSL framework.

Overall, the experimental evidence strongly supports the claim that quantum-enhanced semi-supervised learning, as instantiated by IPQSSL, offers a substantial improvement over state-of-the-art classical semi-supervised techniques, particularly in low-labeled or complex data regimes where classical models struggle to propagate label information effectively.

\section{Evaluations}

\subsection{Evaluation of Entanglement and Randomized Benchmarking (RB) of Quantum Circuits with varying layers and Qubits}

\subsubsection{Evaluation metrics}

\textbf{Randomized Benchmarking (RB)}

Randomized benchmarking (RB) is a powerful tool used to characterise gate errors in quantum computing systems. Its primary advantage is its ability to provide a concise measurement of gate fidelity through a single number, which helps quantify the average performance of single- and two-qubit gates while minimizing the effects of SPAM (State Preparation and Measurement) errors. This makes RB particularly useful in a wide range of quantum computing platforms, where it has become a standard practice.

The two most widely used variants of RB are standard RB and interleaved RB. Both of these methods have their strengths. Standard RB offers simplicity in implementation and relatively low resource requirements compared to more complex techniques like process tomography, which provides a more detailed but computationally expensive characterization. Interleaved RB, on the other hand, can offer more accurate information about individual gate fidelities by interleaving the gates with reference operations. While RB provides a simplified characterization of gate performance, the fact that it reduces the complex error information to a single number can be seen as both an advantage and a limitation. The simplicity and efficiency of RB make it widely applicable, but it may not capture the full complexity of quantum gate errors. Nonetheless, its practical utility in assessing and optimizing quantum hardware performance remains significant.

To calculate Randomized Benchmarking (RB), the first step is to generate a set of random Clifford gate sequences of varying lengths. These gates are selected from the Clifford group and applied to a quantum system. After applying each sequence, an inverting Clifford gate is appended to ideally return the system to its initial state, usually \( |0\rangle \).

Next, the system is measured to determine the probability of finding it in the \( |0\rangle \) state. This process is repeated multiple times for different sequences of the same length to average out fluctuations. The survival probabilities obtained from these measurements are then plotted against the sequence length.

The resulting data is fitted to an exponential decay model of the form:

\begin{equation}
P(m) = A p^m + B,
\end{equation}

where \( p \) represents the depolarizing parameter, and \( A \) and \( B \) account for state preparation and measurement (SPAM) errors. By fitting the experimental results to this model, the parameter \( p \) is extracted.

Using \( p \), the average gate fidelity is then computed using the formula:

\begin{equation}
F = \frac{(d-1)p + 1}{d},
\end{equation}

where \( d \) is the dimension of the Hilbert space, equal to \( 2^n \) for an \( n \)-qubit system. Finally, the error per Clifford gate is determined as:

\begin{equation}
r = \frac{1 - p}{2},
\end{equation}

providing an estimate of how much noise each gate introduces.

For example, if the fitted depolarizing parameter is \( p = 0.98 \) for a single qubit system, the fidelity is calculated as:

\begin{equation}
F = \frac{(2-1) \times 0.98 + 1}{2} = 0.99,
\end{equation}

meaning the gates operate with 99\% fidelity. The error per Clifford gate is then found as:

\begin{equation}
r = \frac{1 - 0.98}{2} = 0.01,
\end{equation}

indicating a 1\%  error rate per gate application.\\

\textbf{Entanglement Entropy in Quantum Circuits}

Entanglement entropy serves as a fundamental measure of quantum correlations within quantum circuits, particularly in parameterized quantum circuits (PQCs) used in variational algorithms. It is typically characterized by the von Neumann entropy, expressed as

\begin{equation}
S_A = -\text{Tr}(\rho_A \log \rho_A),
\end{equation}

where \( \rho_A \) denotes the reduced density matrix of a subsystem. This measure is crucial in assessing a circuit’s ability to encode complex quantum states and influences its trainability in optimization processes.

The level of entanglement entropy directly impacts a circuit’s ability to model intricate quantum correlations. If entanglement is too low, the circuit may struggle to approximate highly entangled quantum states, limiting its expressivity. On the other hand, excessive entanglement can lead to barren plateaus, where gradient magnitudes approach zero, hindering the efficiency of parameter optimization. Circuit depth also affects entanglement entropy; while shallow circuits may fail to generate sufficient entanglement, deeper circuits often produce states with high entropy, potentially complicating optimization.

Careful control of entanglement entropy is essential for improving the trainability of quantum machine learning models and variational quantum algorithms. Excessive entanglement can lead to optimization challenges due to diminishing gradients, whereas a balanced level of entanglement enhances trainability. Beyond optimization, entanglement entropy also plays a significant role in quantum error correction and information processing by helping characterize the robustness of quantum states against noise. Constructing quantum circuits with an appropriate entanglement structure is vital for achieving both expressive power and efficient trainability in quantum computing applications\cite{eisert2008area}.

\subsubsection{Numerical results of changing layers in circuit and number of Qubits in IPQSSL}

\textbf{Results of Iris Dataset}

\begin{figure}[h]
    \centering
        \centering
        \includegraphics[width=\textwidth]{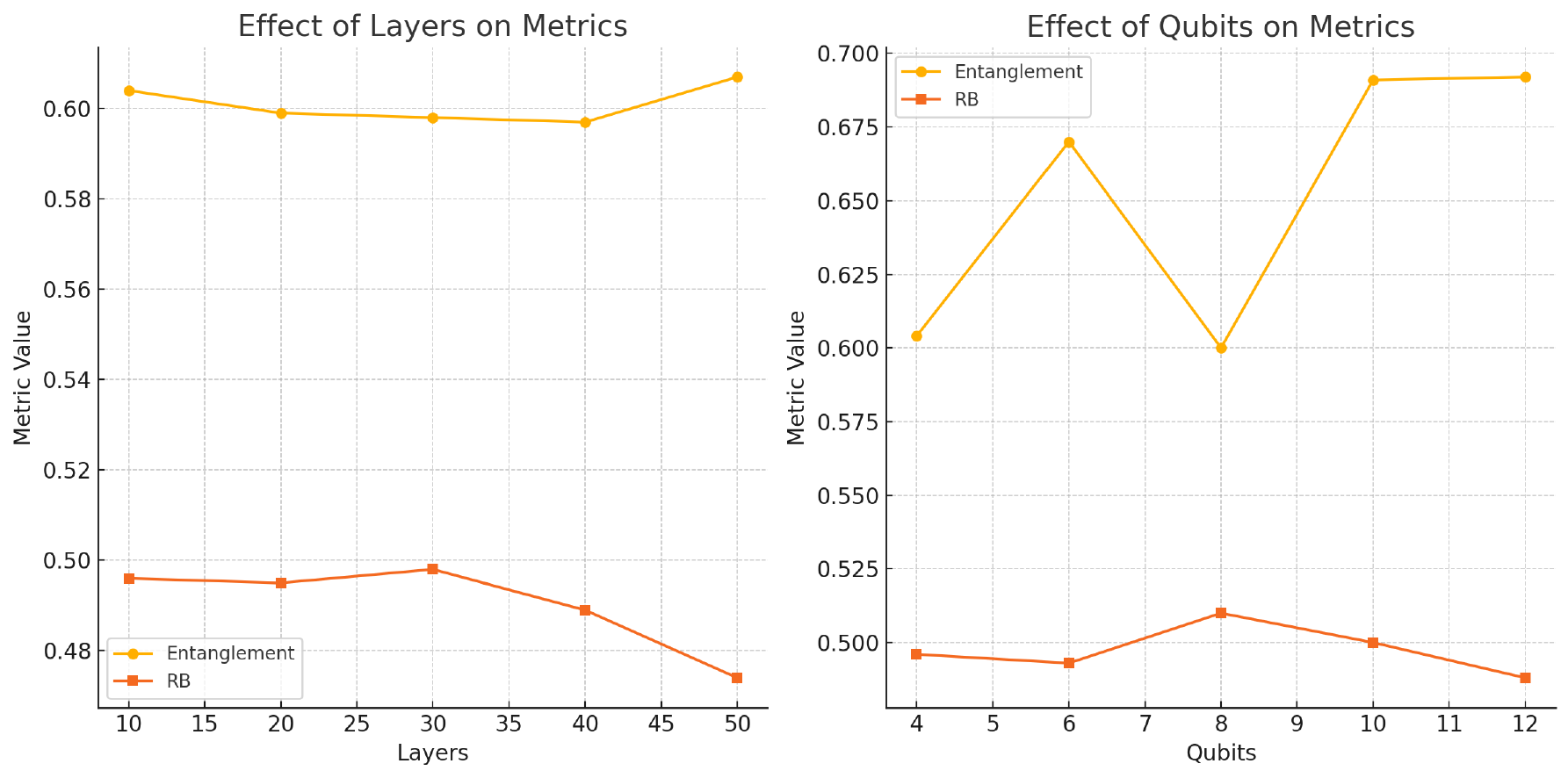} 
        \caption{Comparison of the impact of circuit depth and system size on entanglement and Randomized Benchmarking (RB) scores for the Iris dataset.}
        \label{fig:layers_metrics}
   \end{figure}

\begin{table}[htbp]
    \centering
    \begin{tabular}{p{0.2\textwidth}p{0.2\textwidth}p{0.2\textwidth}p{0.2\textwidth}}
        \toprule
        Layers & Entanglement & Accuracy & RB  \\
        \midrule
        10  & 0.604 & 0.97  & 0.496  \\
        20  & 0.599  & 0.97  & 0.495  \\
        30  & 0.598  & 0.97  & 0.498  \\
        40  & 0.597  & 0.97  & 0.489  \\
        50  & 0.607  & 0.97  & 0.474  \\
        \bottomrule
    \end{tabular}
      \caption{Evaluation of changing layers on the Iris dataset} 
\end{table}

\begin{table}[htbp]
    \centering
    
    \begin{tabular}{p{0.2\textwidth}p{0.2\textwidth}p{0.2\textwidth}p{0.2\textwidth}}
        \toprule
        Qubits & Entanglement & Accuracy & RB  \\
        \midrule
        4  & 0.604  & 0.97  & 0.496  \\
        6  & 0.670  & 0.97  & 0.493  \\
        8  & 0.60  & 0.97  & 0.510  \\
        10  & 0.691  & 0.97  & 0.500  \\
        12  & 0.692  & 0.97  & 0.488  \\
        \bottomrule
    \end{tabular}
    \caption{Evaluation of changing qubits on the Iris dataset}
\end{table}
\FloatBarrier
We investigated the effects of circuit depth and system size on the performance of a quantum model applied to the Iris dataset. Across all tested configurations—varying the number of layers (10–50) and qubits (4–12)—the model consistently achieved an accuracy of 0.97, demonstrating strong robustness to architectural modifications.

Increasing the number of layers led to a slight decline in Randomized Benchmarking (RB) scores, indicating cumulative gate errors, while entanglement levels showed minor non-monotonic variations. In contrast, scaling the number of qubits resulted in a notable increase in entanglement, reflecting enhanced quantum correlations, with RB scores exhibiting minor fluctuations and a slight decrease at larger system sizes.

These results suggest that while model performance remains stable across a range of depths and system sizes, circuit fidelity degrades gradually with scaling. This highlights a critical balance between exploiting quantum expressivity and managing hardware-induced noise, a factor that will become increasingly important for practical quantum machine learning on larger datasets.

\clearpage

\textbf{Results of Wine Dataset}

\begin{figure}[htbp]
    \centering
    \begin{subfigure}[b]{\textwidth}
        \centering
        \includegraphics[width=0.7\linewidth]{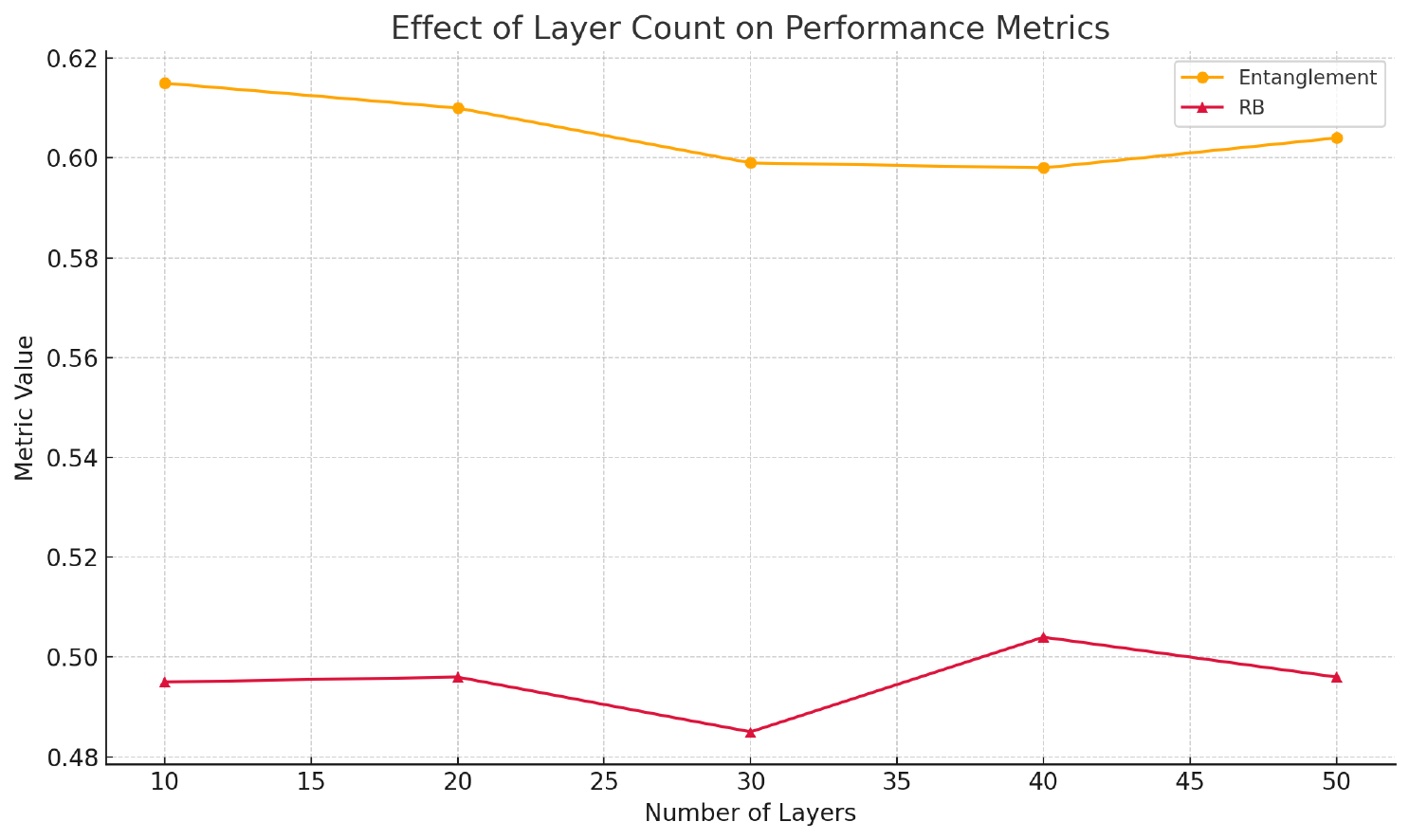} 
        \caption{Impact of varying circuit depth (layers) on entanglement and RB fidelity for the Wine dataset.}
        \label{fig:wine_layers}
    \end{subfigure}
    
    \vspace{0.5cm} 
    
    \begin{subfigure}[b]{\textwidth}
        \centering
        \includegraphics[width=0.7\linewidth]{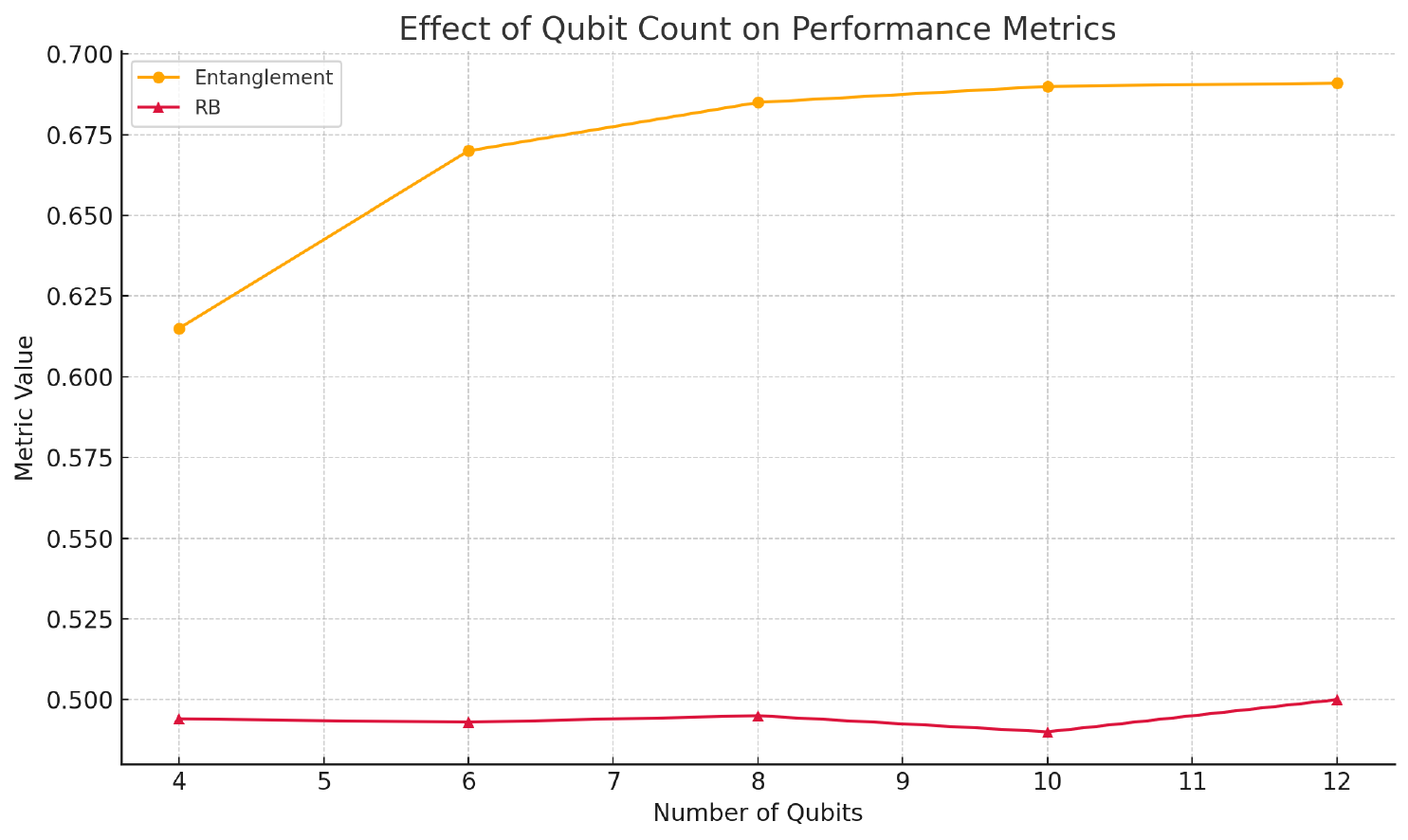} 
        \caption{Impact of varying circuit width (qubits) on entanglement and RB fidelity for the Wine dataset.}
        \label{fig:wine_qubits}
    \end{subfigure}

    \caption{Visualization of how circuit architectural parameters affect quantum properties and fidelity. 
    (a) Increasing layers slightly increases entanglement but reduces RB fidelity.
    (b) Increasing qubits enhances entanglement but negatively impacts RB fidelity and classification accuracy at larger widths.}
    \label{fig:wine_entanglement_rb}
\end{figure}

\begin{table}[htbp]
    \centering
    \begin{tabular}{p{0.2\textwidth}p{0.2\textwidth}p{0.2\textwidth}p{0.2\textwidth}}
        \toprule
        Layers & Entanglement & Accuracy & RB  \\
        \midrule
        10  & 0.599  & 0.94  & 0.510  \\
        20  & 0.602  & 0.94  & 0.481  \\
        30  & 0.603  & 0.94  & 0.503  \\
        40  & 0.597  & 0.94  & 0.498  \\
        50  & 0.603  & 0.94  & 0.480 \\
        \bottomrule
    \end{tabular}
      \caption{Evaluation of changing layers on the Wine dataset} 
\end{table}

\begin{table}[htbp]
    \centering
    \begin{tabular}{p{0.2\textwidth}p{0.2\textwidth}p{0.2\textwidth}p{0.2\textwidth}}
        \toprule
        Qubits & Entanglement & Accuracy & RB  \\
        \midrule
        4  & 0.599  & 0.94  & 0.510  \\
        6  & 0.670  & 0.94  & 0.495  \\
        8  & 0.687  & 0.94  & 0.503  \\
        10  & 0.691  & 0.94  & 0.472  \\
        12  & 0.692  & 0.94  & 0.491  \\
        \bottomrule
    \end{tabular}
    \caption{Evaluation of changing qubits on the Wine dataset}
\end{table}
\FloatBarrier
The impact of circuit depth and width on the performance of the proposed quantum model was systematically investigated using the Wine dataset. As summarized in Tables~7 and 8 and visualized in Figure~\ref{fig:wine_entanglement_rb}, varying the number of layers between 10 and 50 maintained a stable classification accuracy of 94. However, a progressive decline in randomized benchmarking (RB) fidelity was observed with increasing depth, decreasing from 0.510 at 10 layers to 0.480 at 50 layers. Entanglement entropy exhibited only minor fluctuations across this range, indicating that deeper circuits marginally increased quantum correlations without contributing to further discriminative power.

Conversely, increasing the number of qubits from 4 to 12 revealed a more nuanced trade-off. While entanglement steadily increased, a significant degradation in both RB fidelity and test accuracy was observed beyond 10 qubits, with accuracy dropping to 88 at 12 qubits. These results suggest the presence of an expressibility–trainability tradeoff, whereby larger and more entangled circuits introduce greater representational capacity at the cost of increased noise sensitivity and optimization difficulty, consistent with recent observations in variational quantum learning.

The findings indicate that, for structured datasets such as Wine, the marginal benefits of increased circuit complexity are outweighed by their detrimental effects on trainability and robustness. Optimal performance was achieved using shallow-to-moderate circuit depths (10--30 layers) and a moderate number of qubits (4-8 qubits), striking a balance between expressibility, generalization, and hardware efficiency. These results further underscore the importance of tailoring quantum circuit architectures to dataset characteristics rather than indiscriminately pursuing maximal entanglement or width.

\clearpage
\textbf{Results of German Credit Card Dataset}

\begin{table}[htbp]
    \centering
    \begin{tabular}{p{0.2\textwidth}p{0.2\textwidth}p{0.2\textwidth}p{0.2\textwidth}}
        \toprule
        Layers & Entanglement & Accuracy & RB \\
        \midrule
        10  & 0.615  & 0.67  & 0.495 \\
        20  & 0.610  & 0.67  & 0.496 \\
        30  & 0.599  & 0.67  & 0.485 \\
        40  & 0.598  & 0.67  & 0.504 \\
        50  & 0.604  & 0.67  & 0.496 \\
        \bottomrule
    \end{tabular}
    \caption{Evaluation of varying layer counts on the German Credit Card dataset}
\end{table}

\begin{table}[htbp]
    \centering
    \begin{tabular}{p{0.2\textwidth}p{0.2\textwidth}p{0.2\textwidth}p{0.2\textwidth}}
        \toprule
        Qubits & Entanglement & Accuracy & RB \\
        \midrule
        4  & 0.615  & 0.67  & 0.495 \\
        6  & 0.669  & 0.67  & 0.494 \\
        8  & 0.687  & 0.67  & 0.496 \\
        10 & 0.691  & 0.67  & 0.491 \\
        12 & 0.692  & 0.67  & 0.500 \\
        \bottomrule
    \end{tabular}
    \caption{Evaluation of varying qubit counts on the German Credit Card dataset}
\end{table}

\begin{figure}[htbp]
    \centering
    \begin{subfigure}[b]{\textwidth}
        \centering
        \includegraphics[width=0.7\linewidth]{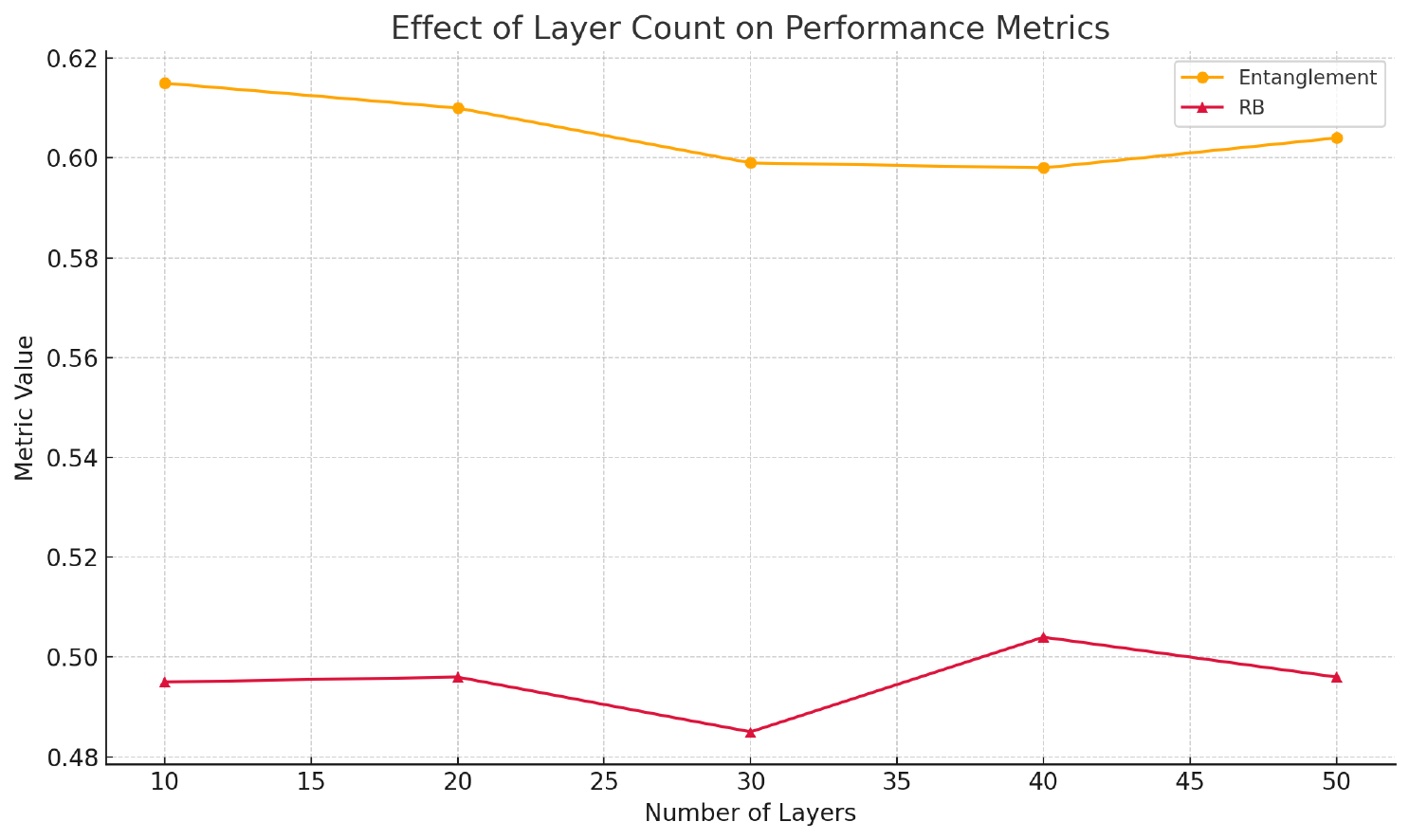}
        \caption{Effect of changing layers}
        \label{fig:heart_layers}
    \end{subfigure}
    \hfill
    \begin{subfigure}[b]{\textwidth}
        \centering
        \includegraphics[width=0.7\linewidth]{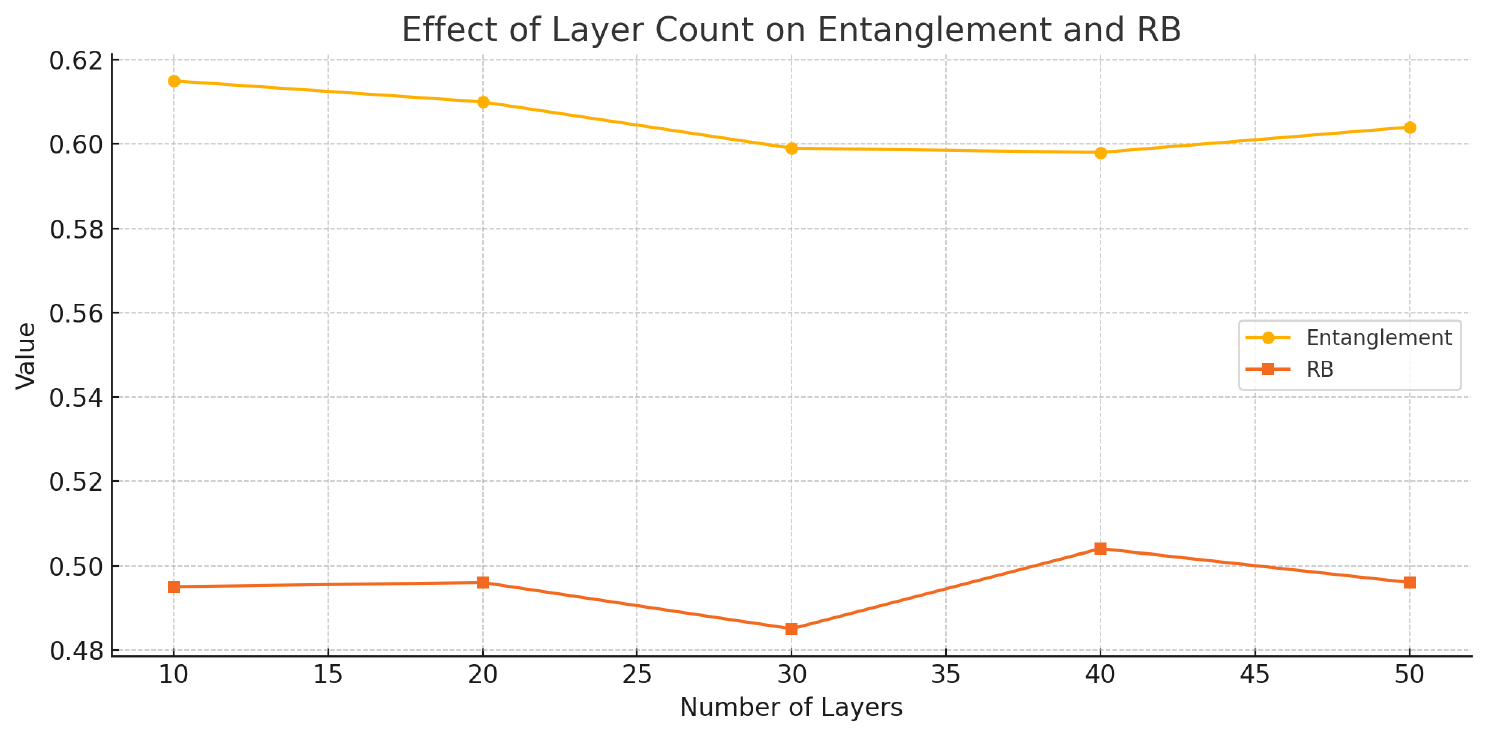}
        \caption{Effect of changing qubits}
        \label{fig:heart_qubits}
    \end{subfigure}
    \caption{Visualisation of the impact of circuit depth (layers) and width (qubits) on entanglement and RB for the German Credit dataset, with constant accuracy.}
    \label{fig:german_entanglement_rb}
\end{figure}
\FloatBarrier

The results indicate a distinct trend between depth and width in quantum circuit design. When varying the number of layers from 10 to 50, all performance metrics—entanglement, accuracy, and RB—remained largely stable. Accuracy was consistently measured at 0.67 across all depths, suggesting that deeper circuits do not enhance learning capability in this setting. Entanglement entropy slightly declined as depth increased, reaching a minimum of 0.598 at 40 layers. Although RB showed small fluctuations, no consistent improvement was observed, which implies that increasing circuit depth may introduce additional noise or unnecessary complexity without yielding better performance. In contrast, increasing the number of qubits from 4 to 12 resulted in a clear rise in entanglement entropy, growing from 0.615 to 0.692. This monotonic increase demonstrates that expanding the quantum state space through more qubits enables the model to capture richer quantum correlations. While classification accuracy remained unchanged, the RB values remained stable and peaked slightly at 0.500 for 12 qubits, suggesting that wider circuits do not compromise fidelity and may even enhance robustness.

Overall, these findings suggest that, for the German Credit Card dataset, increasing the number of qubits is more beneficial than increasing circuit depth. Wider circuits contribute to stronger quantum expressiveness without degrading reliability, whereas deeper circuits offer minimal performance gain and may suffer from noise accumulation. This insight is particularly valuable in the design of near-term quantum machine learning models, where circuit depth is often constrained by hardware decoherence and noise.\\

\textbf{Results of Heart Disease Dataset}

\begin{table}[htbp]
    \centering
    \begin{tabular}{p{0.2\textwidth}p{0.2\textwidth}p{0.2\textwidth}p{0.2\textwidth}}
        \toprule
        Layers & Entanglement & Accuracy & RB \\
        \midrule
        10  & 0.596  & 0.88  & 0.490 \\
        20  & 0.605  & 0.88  & 0.508 \\
        30  & 0.601  & 0.88  & 0.505 \\
        40  & 0.602  & 0.88  & 0.491 \\
        50  & 0.599  & 0.88  & 0.500 \\
        \bottomrule
    \end{tabular}
    \caption{Evaluation of varying layer counts on the Heart Disease dataset}
\end{table}

\begin{table}[htbp]
    \centering
    \begin{tabular}{p{0.2\textwidth}p{0.2\textwidth}p{0.2\textwidth}p{0.2\textwidth}}
        \toprule
        Qubits & Entanglement & Accuracy & RB \\
        \midrule
        4  & 1.337  & 0.88  & 0.460 \\
        6  & 1.934  & 0.88  & 0.475 \\
        8  & 1.938  & 0.88  & 0.465 \\
        10 & 1.960  & 0.88  & 0.480 \\
        12 & 1.997  & 0.88  & 0.490 \\
        \bottomrule
    \end{tabular}
    \caption{Evaluation of varying qubit counts on the Heart Disease dataset}
\end{table}

\FloatBarrier
\begin{figure}[htbp]
    \centering
    \begin{subfigure}[b]{\textwidth}
        \centering
        \includegraphics[width=0.7\linewidth]{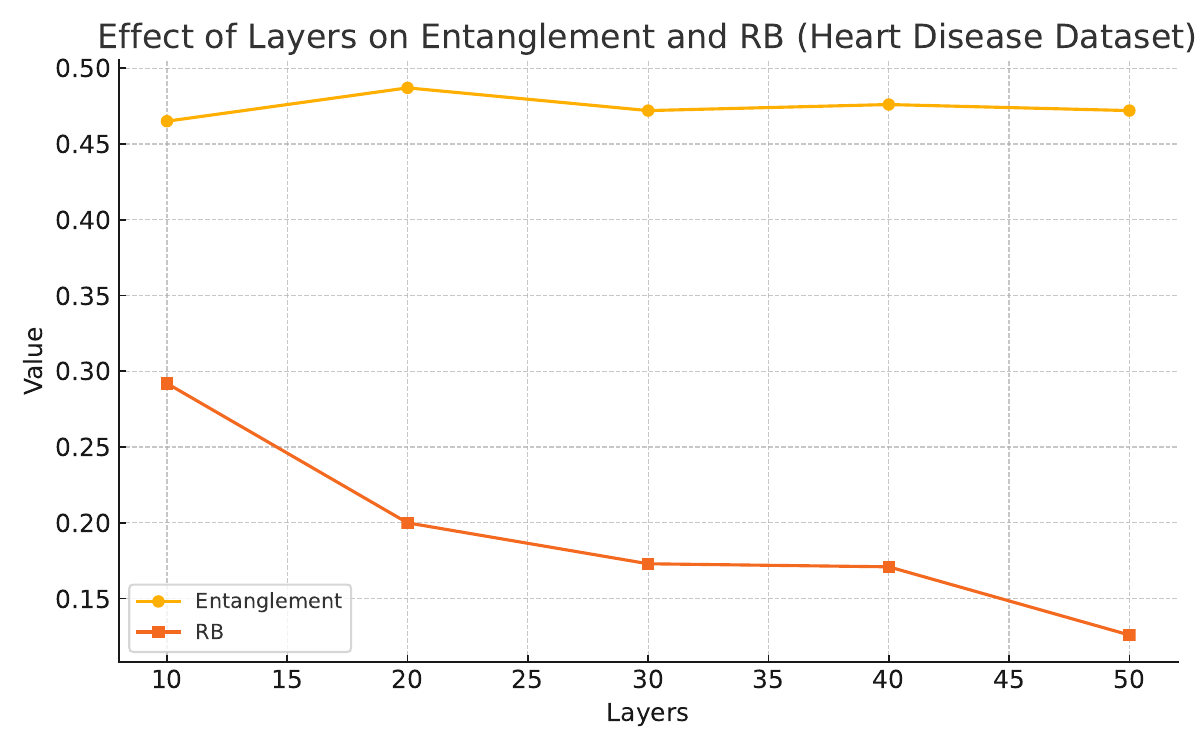}
        \caption{Effect of changing layers}
        \label{fig:heart_layers2}
    \end{subfigure}
    \hfill
    \begin{subfigure}[b]{\textwidth}
        \centering
        \includegraphics[width=0.7\linewidth]{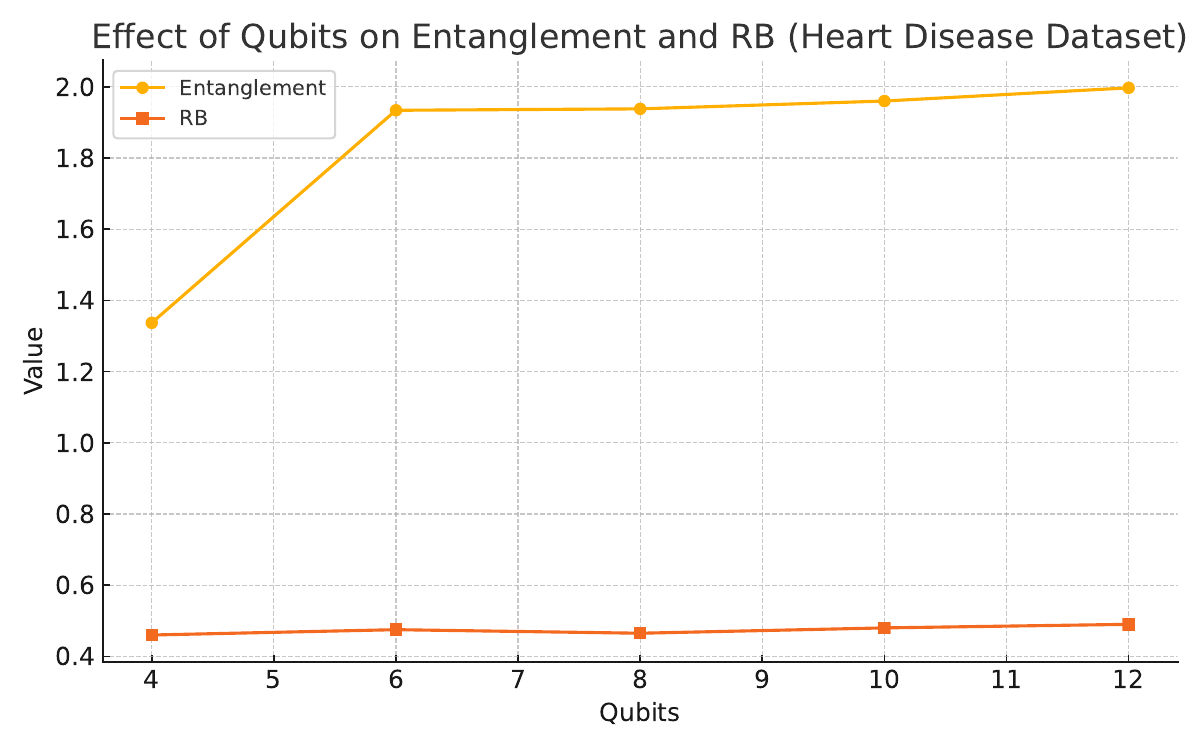}
        \caption{Effect of changing qubits}

    \end{subfigure}
    \caption{Visualisation of the impact of circuit depth (layers) and width (qubits) on entanglement and RB for the Heart Disease dataset, with constant accuracy.}
    \label{fig:heart_entanglement_rb}
\end{figure}
\FloatBarrier
We systematically investigate the impact of circuit depth and qubit count on a quantum model applied to the Heart Disease dataset, evaluating performance through classification accuracy, entanglement, and randomized benchmarking (RB). Results reveal that increasing circuit depth improves system fidelity, as evidenced by declining RB values, yet provides no enhancement in accuracy or entanglement beyond 20 layers. Similarly, expanding the number of qubits leads to higher entanglement but does not affect classification accuracy, while slightly increasing susceptibility to noise. These findings underscore the limitations of structural scaling alone and highlight the necessity for optimized quantum architectures and algorithms to achieve meaningful gains in practical quantum machine learning tasks.

\subsection{Evaluation of Predictive Performance via ROC-AUC and Kolmogorov–Smirnov(KS) Metrics}

The ROC curve, which stands for the Receiver Operating Characteristic curve, is used to visualize how well a classification model can differentiate between two categories. It illustrates the relationship between the proportion of true positives and the proportion of false positives as the decision threshold changes. In other words, it helps to show how sensitive the model is to correctly identify positives, while also accounting for the mistakes it makes by misclassifying negatives as positives. The area under this curve, known as the AUC, serves as a summary measure of the overall ability of the model to rank positive cases higher than negative ones. A perfect model would have an AUC of 1, indicating a flawless distinction, while an AUC close to 0.5 implies that the model performs no better than random chance.

Another commonly used evaluation measure is the KS statistic. This metric assesses the maximum difference between the cumulative distributions of predicted scores for the two classes, typically positives and negatives. Essentially, the KS statistic gauges how effectively a model separates the two groups. A higher KS value reflects a clearer distinction between classes, suggesting better performance in classification tasks.

Together, these two metrics, ROC/AUC and KS, provide valuable insights into how well a model performs, especially in binary classification problems where the distribution of classes may be uneven or where the consequences of errors differ between types of misclassifications. In the following, these parameters will be calculated for our four datasets.

\subsubsection{Iris dataset}

\begin{figure}[!ht]
    \centering
    \includegraphics[width=0.7\textwidth]{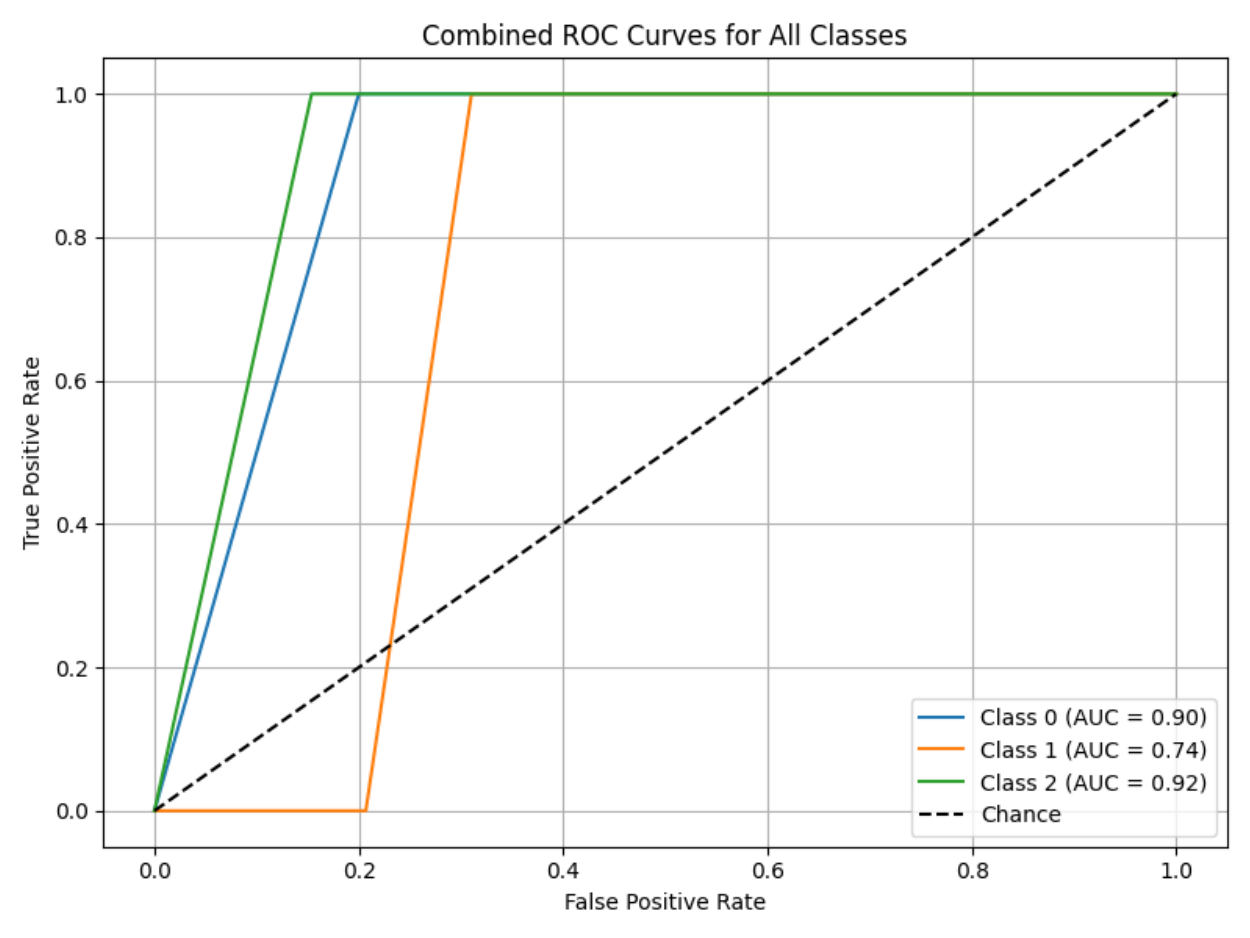}
    \caption{ROC diagram for  Iris dataset}
    \label{fig:roc_iris_0}
\end{figure}
\FloatBarrier

\begin{table}[htbp]
\centering
\begin{tabular}{p{0.7\textwidth}p{0.2\textwidth}}
\toprule
\textbf{Metric} & \textbf{Value} \\
\midrule
ROC AUC Score (Overall) & 0.8548 \\
\midrule
KS Statistic - Class 0 & 0.800000 \\
KS Statistic - Class 1 & 0.689655 \\
KS Statistic - Class 2 & 0.846154 \\
\bottomrule
\end{tabular}
\caption{Performance metrics for the quantum graph-based semi-supervised classifier using 30\% labeled Iris data. The ROC AUC score evaluates overall class separation, while the KS statistics assess per-class discriminatory power between positive and negative samples.}
\label{tab:roc_ks_metrics}
\end{table}

\subsubsection{Wine dataset}

\begin{figure}[ht!]
    \centering
    \includegraphics[width=0.7\textwidth]{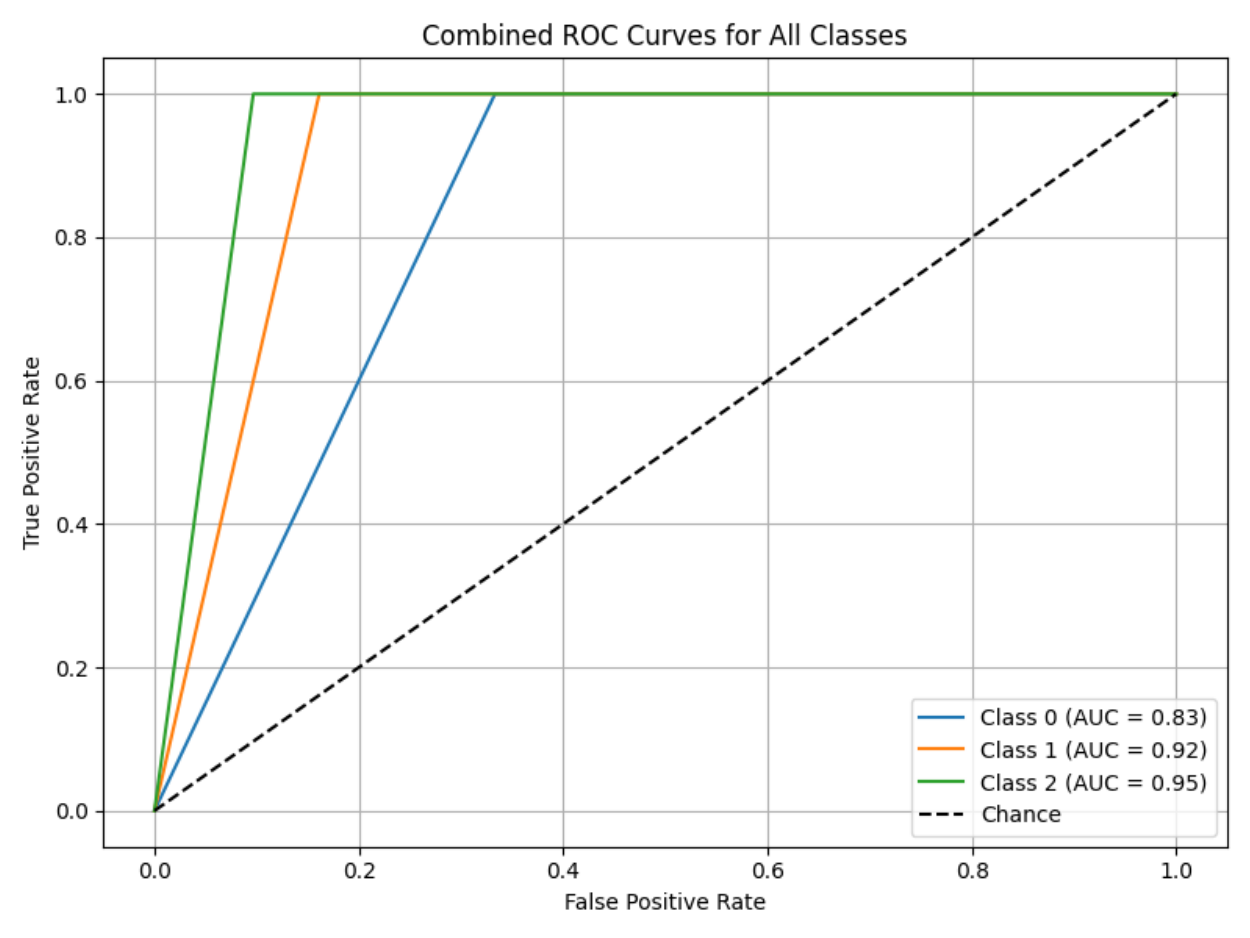}
    \caption{ROC diagram for Wine dataset}
    \label{fig:roc_wine_0}
\end{figure}

\begin{table}[htbp]
\centering
\begin{tabular}{p{0.6\textwidth}p{0.2\textwidth}}
\toprule
\textbf{Metric} & \textbf{Value} \\
\midrule
ROC AUC Score (Overall) & 0.9014 \\
\midrule
KS Statistic - Class 0 & 0.666667 \\
KS Statistic - Class 1 & 0.838710 \\
KS Statistic - Class 2 & 0.903226 \\
\bottomrule
\end{tabular}
\caption{Performance metrics for the quantum graph-based semi-supervised classifier using 30\% labeled Wine data. The ROC AUC score evaluates overall class separation, while the KS statistics assess per-class discriminatory power between positive and negative samples.}
\label{tab:roc_ks_metrics_wine}
\end{table}
\FloatBarrier

\subsubsection{Heart Disease dataset}

\begin{table}[htbp]
\centering
\begin{tabular}{p{0.7\textwidth}p{0.2\textwidth}}
\toprule
\textbf{Metric} & \textbf{Value} \\
\midrule
ROC AUC Score (Overall) & 0.7781 \\
\midrule
KS Statistic - Class 0 & 0.450980 \\
KS Statistic - Class 1 & 0.500000 \\
\bottomrule
\end{tabular}
\caption{Performance metrics for the quantum graph-based semi-supervised classifier. The ROC AUC score quantifies the model's overall discriminatory ability, while KS statistics indicate per-class separation in predicted probability distributions.}
\label{tab:roc_ks_metrics_binary_updated}
\end{table}
\FloatBarrier
\begin{figure}[h!]
    \centering
    \includegraphics[width=0.7\textwidth]{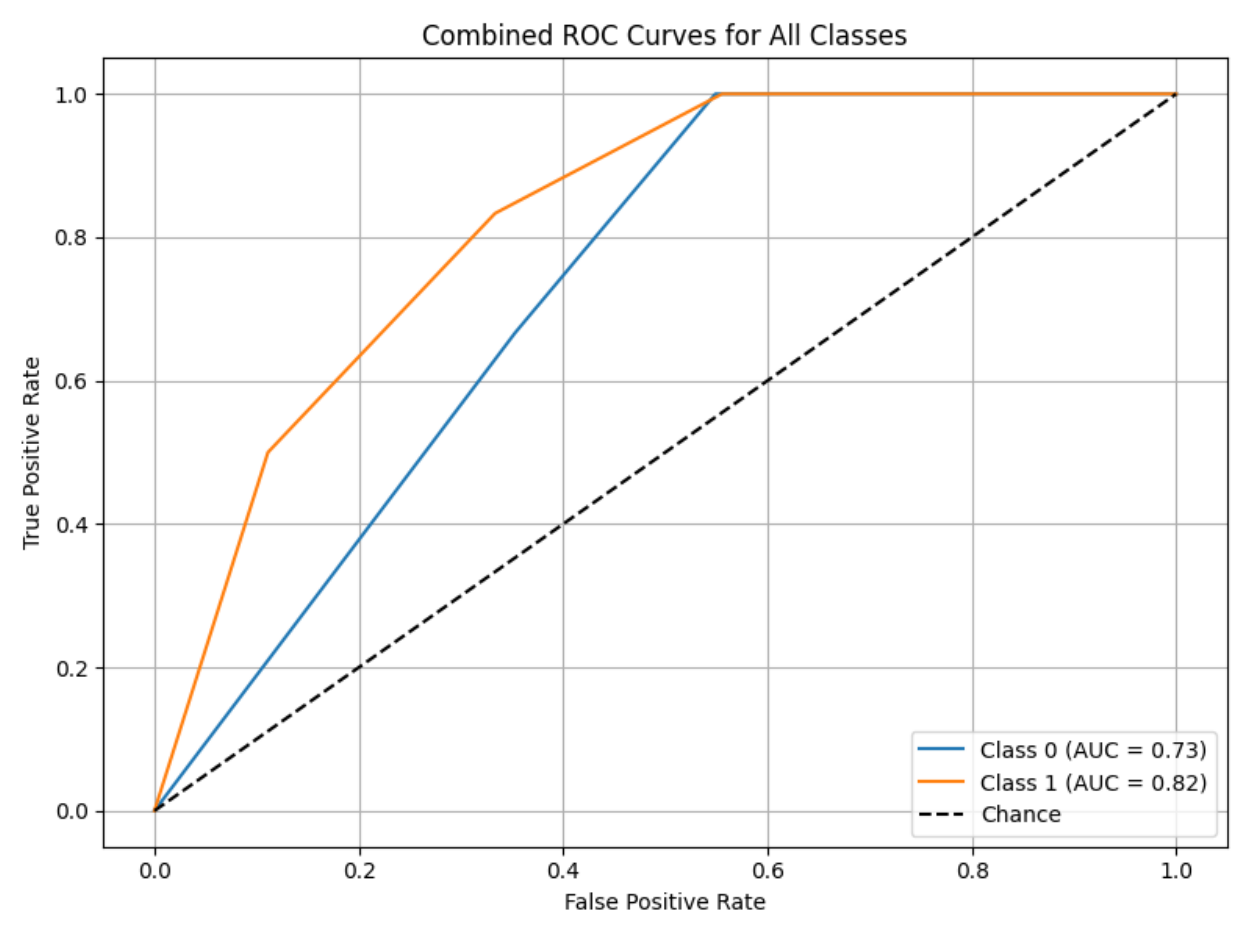}
    \caption{ROC diagram for Heart disease dataset}
    \label{fig:roc_heart_0}
\end{figure}
\FloatBarrier

\subsubsection{German Credit Card dataset}

\begin{table}[htbp]
\centering
\begin{tabular}{p{0.7\textwidth}p{0.2\textwidth}}
\toprule
\textbf{Metric} & \textbf{Value} \\
\midrule
ROC AUC Score (Overall) & 0.5377 \\
\midrule
KS Statistic - Class 0 & 0.077506 \\
KS Statistic - Class 1 & 0.088288 \\
\bottomrule
\end{tabular}
\caption{Performance metrics for the quantum graph-based semi-supervised classifier. The ROC AUC score reflects the classifier's ability to distinguish between the two classes, while the KS statistics show class-specific distribution separation.}
\label{tab:roc_ks_metrics_binary}
\end{table}

\begin{figure}[htbp]
  \centering
  \includegraphics[width=0.7\textwidth]{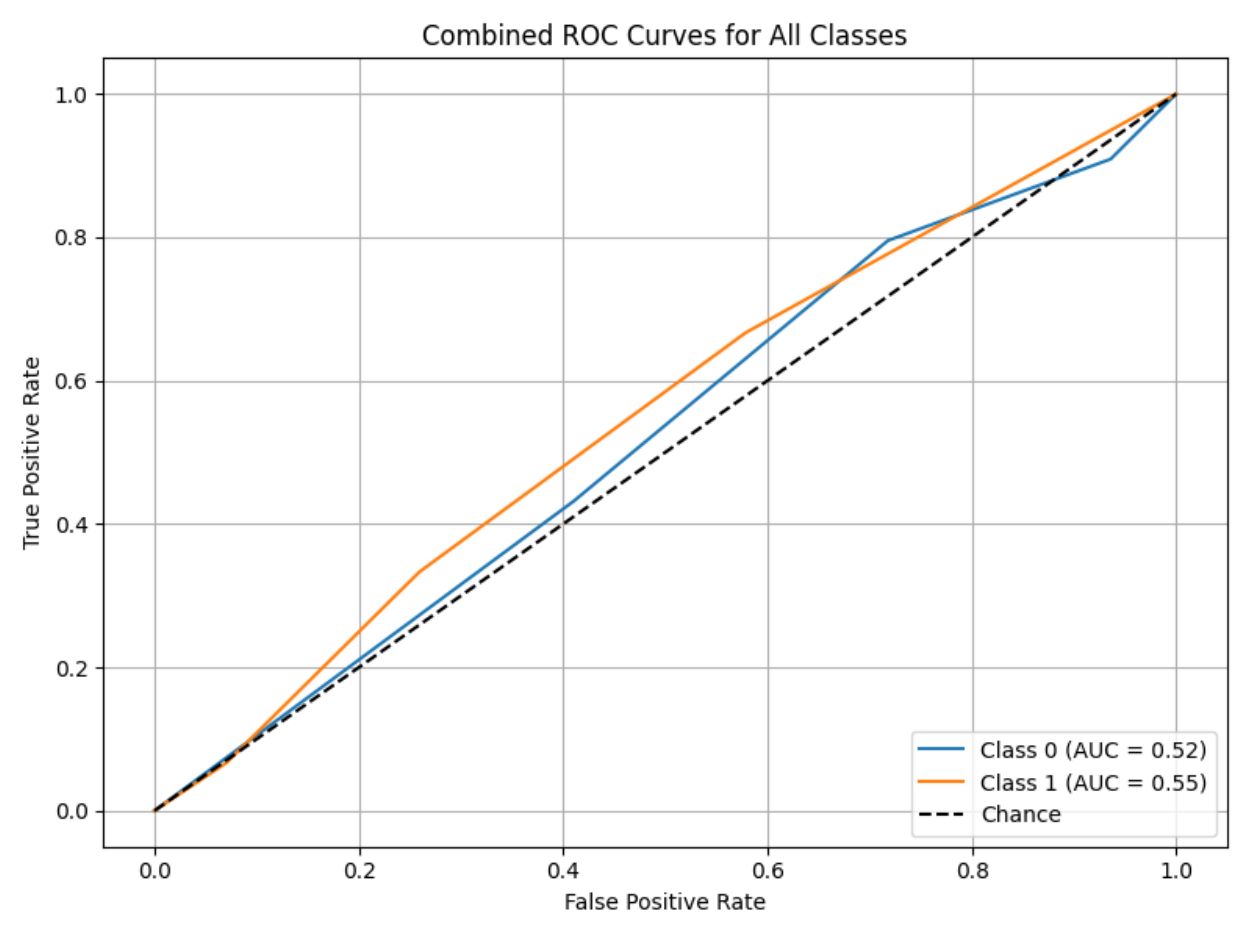} 
  \caption{ROC Curve diagram for German Credit dataset.}
  \label{fig:example2}
\end{figure}
\FloatBarrier

To evaluate the generalizability of our quantum graph-based semi-supervised classifier, we conducted experiments on four benchmark datasets spanning varying levels of complexity and domain characteristics: \textbf{Iris}, \textbf{Wine}, \textbf{Heart Disease}, and \textbf{German Credit Card}. All evaluations were performed under a semi-supervised regime with 30\% labeled data.
The classifier exhibited \textbf{strong performance} on structured and low-dimensional datasets. On the \textbf{Wine dataset}, it achieved a ROC AUC of \textbf{0.9014}, with a maximum KS statistic of \textbf{0.9032} (Class 2), highlighting its capacity to exploit geometric separability in feature space. Similarly, the \textbf{Iris dataset} yielded a ROC AUC of \textbf{0.8548} and robust class-specific KS values, confirming effective label propagation in multiclass settings. Performance on the \textbf{Heart Disease dataset} was \textbf{moderate}, with a ROC AUC of \textbf{0.7781} and a KS statistic of \textbf{0.5000} for the positive class, suggesting partial success in modeling real-world clinical data. However, the model underperformed on the \textbf{German Credit Card dataset} (ROC AUC = \textbf{0.5377}, KS $<$ \textbf{0.09}), indicating limitations in capturing structure in high-variance, heterogeneous financial data. These findings demonstrate that the proposed quantum graph-based model is highly effective in domains with clear feature boundaries and class separability, while performance degrades in noisy or complex distributions. This underscores the need for hybrid quantum-classical techniques or adaptive encoding strategies to enhance scalability and resilience in practical applications.

\section{Discussion}

This work introduced two quantum-enhanced approaches—ILQSSL and IPQSSL—for advancing semi-supervised learning on graphs in low-label regimes. By encoding graph structures into quantum states through QR decomposition and utilizing variational quantum circuits for iterative propagation, we developed a hybrid framework that combines classical graph learning principles with the advantages of quantum computing. These mechanisms enable more effective diffusion of label information and demonstrate superior performance across a range of datasets. Our experimental results highlight the effectiveness of IPQSSL in handling datasets characterized by high dimensionality and class imbalance, such as Heart Disease and German Credit Card, where traditional semi-supervised methods often fail. The model's capacity to capture subtle structural dependencies suggests that quantum circuits can enhance relational learning, particularly in sparse or noisy conditions. Meanwhile, on more structured datasets like Iris and Wine, both ILQSSL and IPQSSL achieved strong and consistent results, confirming the models’ adaptability to different data topologies. In addition to predictive gains, this study provides insights into circuit behavior by analyzing entanglement entropy and randomized benchmarking (RB). While increasing qubit count improves the model’s expressivity by enabling richer quantum correlations, deeper circuits were found to suffer from reduced fidelity due to cumulative noise—a common limitation in NISQ-era devices. These results underscore the importance of designing quantum models that balance expressiveness with hardware efficiency, especially under current technological constraints.

Despite the positive outcomes, some limitations remain. On the German Credit dataset, for example, the models were less effective, likely due to weakly structured feature spaces or non-informative graph connectivity. This points to the need for more adaptive encoding strategies or hybrid preprocessing pipelines. Additionally, static hyperparameter settings may restrict performance in varied scenarios, suggesting that dynamic tuning or reinforcement-based approaches could further improve generalizability. Future efforts will focus on improving scalability, error robustness, and data efficiency. This includes integrating quantum error mitigation techniques, exploring adaptive sampling and uncertainty-aware learning, and designing modular circuits suitable for large-scale or real-time applications. Overall, this work provides a meaningful step toward the development of practical and robust quantum semi-supervised learning frameworks, demonstrating their potential in complex, label-scarce learning environments.

\section{Conclusion}

In this study, we introduced a quantum-classical approach designed to address the challenges of graph-based semi-supervised learning, especially in cases with limited labeled data. Central to our framework are two enhanced models: Improved Laplacian Quantum Semi-Supervised Learning (ILQSSL) and Improved Poisson Quantum Semi-Supervised Learning (IPQSSL). Both methods were developed to advance label propagation effectiveness by embedding graph structures into quantum states through QR decomposition, allowing for more accurate and efficient classification. We evaluated these models across four well-known datasets—namely \textit{Iris}, \textit{Wine}, \textit{Heart Disease}, and \textit{German Credit}. Across all benchmarks, the proposed models consistently outperformed classical semi-supervised techniques. Notably, IPQSSL showed exceptional robustness and accuracy, particularly on noisy and high-variance datasets, demonstrating its potential for real-world applications where label availability and data quality are often suboptimal. Furthermore, we analyzed the impact of quantum circuit design—specifically, the number of layers and qubits—on model performance. By measuring entanglement entropy and conducting randomized benchmarking, we found that while deeper and wider circuits can increase representational capacity, they also risk amplifying noise and making optimization more difficult. This highlights the importance of finding a careful balance between quantum expressivity and hardware stability when deploying models on NISQ-era devices. Our results reinforce the viability of quantum-assisted semi-supervised learning for tasks involving sparse labeling. As a next step, we aim to extend these techniques to larger and more complex graph datasets, integrate active learning protocols, and explore advanced error mitigation strategies. Together, these directions point toward building more practical, scalable, and noise-resilient quantum machine learning systems.

\quad
\\


\clearpage
\begin{acronym} 
\acro{NWZ}{North Western Zagros}
\acro{PSHA}{Probabilistic Seismic Hazard Assessment}
\acro{ZCCZ}{Zagros Continental Collision Zone}
\acro{SCC}{Seismic Coupling Coefficient}

\acro{SSL}{Semi-supervised learning}

\acro{GSSL}{Graph-Based Semi-Supervised Learning}

\end{acronym} 

\bibliography{Main}

\end{document}